\documentstyle[epsfig]{mn}{}
\begin{document}

\def\mpc{h^{-1} {\rm{Mpc}}}
\def\up{h^{-3} {\rm{Mpc^3}}}
\def\uk{h {\rm{Mpc^{-1}}}}
\def\lsim{\mathrel{\hbox{\rlap{\hbox{\lower4pt\hbox{$\sim$}}}\hbox{$<$}}}}
\def\gsim{\mathrel{\hbox{\rlap{\hbox{\lower4pt\hbox{$\sim$}}}\hbox{$>$}}}}
\def\kms {\rm{km~s^{-1}}}
\def\apj {ApJ}
\def\aj {AJ}
\def\aa {A \& A}
\def\mnras {MNRAS}

\title{Galaxy groups in the 2dF galaxy redshift survey: Luminosity and Mass Statistics}
\author[H.J. Mart\'\i nez et al]{
\parbox[t]{\textwidth}{
H.J. Mart\'{\i}nez, 
A. Zandivarez, 
M.E. Merch\'an \&
M.J.L. Dom\'{\i}nguez}
\vspace*{6pt}\\ 
\parbox[t]{15 cm}{
Grupo de Investigaciones en Astronom\'{\i}a Te\'orica y Experimental, 
IATE, Observatorio Astron\'omico, Laprida 854, 5000, C\'ordoba, Argentina} \\
}
\date{\today}

\maketitle

\begin{abstract}
Several statistics are applied to groups and galaxies in groups in the 
Two degree Field Galaxy Redshift Survey.
Firstly we estimate the luminosity functions for different
subsets of galaxies in groups.
The results are well fitted by a Schechter function with parameters 
$M^{\ast}-5\log(h)=-19.90 \pm 0.03$ and $\alpha=-1.13\pm0.02$ for all 
galaxies in groups, which is quite consistent with the results by Norberg
et al. for field galaxies. 
When considering the four different spectral types
defined by Madgwick et al. we find that the characteristic magnitude 
is typically brighter than in the field. We also observe
a steeper value, $\alpha=-0.76\pm 0.03$, 
of the faint end slope for low star-forming galaxies when
compared with the corresponding field value. 
This steepening is more conspicuous, $\alpha=-1.10\pm 0.06$, 
for those galaxies in more massive groups (${\mathcal M}
\gsim 10^{14} h^{-1} M_{\odot}$) than the obtained in the lower mass subset, 
$\alpha=-0.71\pm 0.04$ (${\mathcal M}<10^{14} h^{-1} M_{\odot}$). 

Secondly, we compute group total luminosities using Moore, Frenk
\& White prescriptions.
We define a flux-limited group sample using a new statistical tool
developed by Rauzy.
The resulting group sample is used to determine the group luminosity function
finding a good agreement with previous determinations and semianalytical 
models.

Finally, the group mass function for the flux-limited sample is derived.
An excellent agreement is obtained when comparing our determination with
analytical predictions over two orders of magnitude in mass.

\end{abstract}


\section{Introduction} 
Groups of galaxies constitute one of the most suitable laboratories
for the study of properties of intermediate galaxy density environments
and their consequences on the process of galaxy formation and evolution.
Furthermore, several hints about the large scale structure of the universe
and how structures evolve in the universe can be drawn from the statistical
studies of groups and their properties.
Some of them, for instance the luminosities and morphological types of 
their member galaxies, are sensitive to the processes of mergers and 
interactions between individual galaxies inside a potential well.
Meanwhile, other properties as group abundance as a function of total
luminosity or mass, can provide constraints to hierarchical clustering 
scenarios and cosmological models.  
In this paper, we provide a robust statistical study about
the luminosities of group galaxy members, and 
also on global group properties such as
total luminosities and masses, using one of the largest 
group catalogues at the present. The accuracy of these determinations
allow us a fair comparison with analytical and semianalytical predictions.

Most studies on the environmental dependence of the galaxy luminosity function
have been carried out either in the field or in rich clusters using 
the largest two dimensional catalogues and small redshift surveys.
Regarding field galaxy luminosity functions, several works have been devoted
to its determination in the last decade (Loveday et al. 1992, Marzke, Huchra
\& Geller 1994, Lin et al. 1996, Zucca et al. 1997, Ratcliffe et al. 1998).
With the advent of large redshift surveys such as the Two degree Field Galaxy 
Redshift Survey (2dFGRS) and the Sloan Digital Sky Survey (SDSS), more reliable 
statistical results have been obtained. The field luminosity functions
determined by Blanton et al. (2001) (SDSS) and Norberg et al. (2002) (2dFGRS)
show an excellent agreement, finding a luminosity function accurately
described by a Schechter function with parameters $M_{b_J}^{\ast}-5\log(h)
\simeq-19.66$ and $\alpha\simeq-1.21$. In particular, the 2dFGRS allowed
the determination of the field luminosity function for galaxies of different
spectral types (Madgwick et al 2002) finding a systematic steepening of
the faint end slope moving from passive $(\alpha=-0.54)$ to active 
$(\alpha=-1.5)$ star forming galaxies, and also a corresponding faintening
of $M^{\ast}$. 

A controversial issue about the luminosity function of galaxies in clusters
is the steepening at the faint end, compared with field galaxy luminosity
function (Valotto et al 1997, Trentham 1997, L\'opez-Cruz et al 1997).    
Recently Goto et al. (2002) computed a composed luminosity
function using 204 clusters taken from SDSS (York et al. 2000). They found 
that the slopes of the LF's become flatter toward redder color band and that
have brighter characteristic magnitude and flatter slopes than the field LF.

The lack of statistical results on smaller overdensities
such as groups of galaxies is mainly due to the fact that
two dimensional identification privileges the
largest overdensities. Analyzing a sample of 66 groups of galaxies
identified in redshift space, 
Muriel, Valotto \& Lambas (1998) find a flat faint end for the
galaxy luminosity function in groups ($\alpha \simeq -1.0$) 
compared with the luminosity function in clusters where a 
large relative number of faint galaxies is present.
It is important to remark that most of the luminosity
function estimations in groups and clusters of galaxies are computed 
subtracting background and foreground contamination due to the lack
of spectroscopic information for galaxy members.

One step further, in order to understand the transition between galaxy
and galaxy systems luminosities, is the computation of the luminosity function 
of galaxy groups. Moore, Frenk \& White (1993), reanalysing the groups in the
Center for Astrophysics (CfA) redshift survey, developed a method for the
estimation
of the total luminosity of groups identified in magnitude-limited galaxy 
surveys. This method allowed them the computation of the luminosity function of 
galaxy systems for a sample of 163 groups with at least tree members. 
Another attempt to determine the luminosity function of virialized systems
was made by Marinoni, Hudson \& Giuricin (2002). They used the Nearby Optical
Galaxy (NOG) sample, which comprise $\sim 7000$ galaxies with $cz \leq 6000 \kms
$ and $B \leq 14$, finding a very good agreement with Moore, Frenk \&
White (1993) previous determination. 

On the other hand, the abundance of haloes as a function of mass constitutes
a key point in both, the determination of a cosmological model and the
understanding of the structure collapse. 
At the present, the more popular models for halo abundance are the analytical
model of Press \& Schechter (1974) for spherical collapse, the Sheth \&
Tormen (1999) model for ellipsoidal collapse and Jenkins et al. (2001) 
fit obtained from numerical simulations.
Many efforts have been made to determine the mass function of galaxy systems
from observations (Bahcall \& Cen 1993, Biviano et al. 1993, Girardi et al.
1998). Recently, Girardi \& Giuricin (2000) have computed the mass function
for a sample of nearby loose groups by Garcia (1993). 
They found the group mass function to be a smooth extrapolation of the 
cluster mass function and a reasonable agreement with the Press \& Schechter
(1974) predictions.

Currently, one of the largest group catalogue was constructed by
Merch\'an \& Zandivarez (2002).
They have identified groups on the 2dF public 
100K data release using a modified Huchra \& Geller (1982) group 
finding algorithm that takes into account 2dF magnitude limit and 
redshift completeness masks.
This catalogue constitutes a large and suitable sample for both, the study
of processes in group environment and the properties of the group population 
itself.
The global effects of group environment on star formation
was analysed by Mart\'{\i}nez et al. (2002) using this catalogue.
They have found a strong correlation
between the relative fraction of different galaxy types and the parent group
virial mass. For groups with $M\gsim 10^{13} M_{\odot}$ the relative fraction
of star forming galaxies is significantly suppressed, indicating that
even intermediate mass environments affect star formation.
Dom\'{\i}nguez et al. (2002) presented hints toward understanding local  
environment effects affecting the spectral types of galaxies in groups
by studying the relative fractions of different spectral types
as a function of the projected local galaxy density and
the group-centric distance. A similar analysis were performed in known
galaxy clusters and their environments in the 2dFGRS by Lewis et al (2002).

The aim of this work is to use the Merch\'an \& Zandivarez (2002) 
group catalogue to obtain reliable
determinations of internal and global properties of groups:
luminosity functions of galaxies in groups, group luminosity and mass
functions. 
The outline of this paper is as follows. In section 2, we present a revised
version of the 2dF Galaxy Group Catalogue (2dFGGC) used throughout this work. 
Section 3 describes the methods and results of the luminosity function of 
galaxies in groups while in section 4 description corresponds to the luminosity 
function of galaxy groups. 
The computation of group mass function and a comparison with analytical 
models are presented in section 5. Finally, in section 6 we summarize our
conclusions.

\section{The 2dFGGC}

Samples of galaxies and groups used in this work are a new sample
constructed using a revised version of the masks and mask software of the 
2dFGRS 100k data release, which includes the $\mu$-masks described in 
Colless et al. (2001) 
(see http://msowww.anu.edu.au/2dFGRS/Public/Release/
Masks/index.html).
The group catalogue is obtained following the same procedure as described 
by Merch\'an \& Zandivarez (2002). In the previous work, 
the finder algorithm used for group identification is similar to that 
developed by Huchra \& Geller (1982) but modified in order to take into account 
redshift completeness and magnitude limit mask
present on the current release of galaxies. Here we include
the effect introduced by the magnitude completeness mask ($\mu$-mask).
In the construction of the 2dFGGC values of $\delta \rho/\rho=80$ 
and $V_0=200~\kms$ were used to maximize the group accuracy. 
The revised group catalogue comprises a total number
of 2198 galaxy groups with at least 4 members and
mean radial velocities in the range $900 ~\kms\leq V \leq 75000~\kms$.
These groups have a mean velocity dispersion of 
$265~ \kms$, a mean virial mass of $9.1\times 10^{13} \ h^{-1} \ 
M_{\odot}$ and a mean virial radius of $1.15 ~ \mpc$.
These results show that the new identification keeps the mean properties
of the group catalogue obtained by Merch\'an \& Zandivarez (2002), as
expected from the fact that the introduction of the magnitude completeness
mask is a second order correction.

\section{Luminosity function of galaxies in groups}

\begin{figure}
\epsfxsize=0.5\textwidth
\hspace*{-0.5cm} \centerline{\epsffile{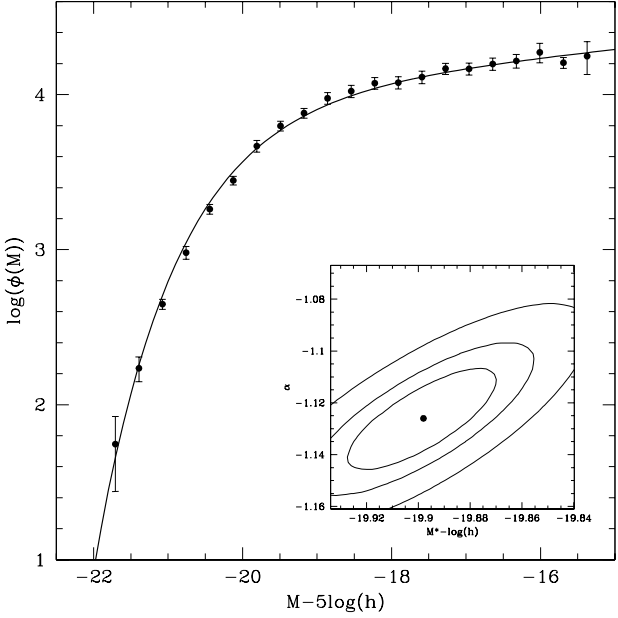}}
\caption{ 
$b_J$ band luminosity function of galaxies in groups from the
2dFGGC (arbitrary units) computed using the $C^-$ method.
Error bars were estimated using mock catalogues as described in text.
Solid line shows the STY Schechter fit to the data.
Inset panel displays $1\sigma, 2\sigma$ and $3\sigma$ confidence ellipses
enclosing the best fit Schechter function parameters $\alpha=-1.13\pm 0.02$ 
and $M^{\ast}-5\log h=-19.90\pm 0.03$.
}
\label{fig1}
\end{figure}

In this section we estimate the luminosity functions (LF) of galaxies in
groups. The following analysis comprises the study of the whole sample
of galaxies in groups and different subsets defined by galaxy spectral types
as defined by Madgwick et al (2002). 
This classification is based on the $\eta$ parameter which is very tightly
correlated with the equivalent width of $H_{\alpha}$ emission line,  
correlates well with morphology for emission line galaxies and can be interpreted
as a measure of the relative current star-formation present in each galaxy.
The four spectral types are defined as:
\begin{itemize}
\item Type 1: $~~~~~~~~~~\eta < -1.4$,
\item Type 2: $-1.4\leq \eta < ~~1.1$,
\item Type 3: $~~1.1 \leq \eta < ~~3.5$, 
\item Type 4: $~~~~~~~~~~\eta\ge ~~3.5$.
\end{itemize}
The Type 1 class is characterised by an old stellar population and
strong absorption features, the Types 2 and 3 comprise spiral
galaxies with increasing star formation, finally the Type 4
class is dominated by particularly active galaxies such as starburst.

\subsection{Luminosity function estimators}

In a comparative study of different LF estimators using Monte Carlo simulations,
Willmer (1997) shows that the $C^-$ method of Lynden-Bell (1971) and the STY
method derived by Sandage, Tammann \& Yahil (1979) are the best estimators
to measure the shape of the LF. Moreover, Willmer's analysis states that
the $C^-$ is the most robust 
estimator, being less affected by different values of the faint end slope
of the Schechter parametrisation and sample size.
In this work we use both, the $C^-$ method to make a non parametric 
determination of the LF and the STY method to calculate the maximum likelihood
Schechter fit to the LF.

The $C^-$ method was simplified and developed by 
Choloniewski (1987) in order to compute
simultaneously the shape and normalisation of the luminosity function.
The LF is obtained by differentiating the cumulative LF, $\Psi(M)$. 
The function
$X(M)$ defined as the observed density of galaxies with absolute 
magnitude brighter than $M$, represents only an undersampling of the $\Psi(M)$, 
\begin{equation}
\frac{dX}{X}<\frac{d\Psi}{\Psi};
\end{equation}
the key point of the method is to construct a quantity $C(M)$, subsample
of $X(M)$, such that 
\begin{equation}
\frac{dX}{C}=\frac{d\Psi}{\Psi}.
\end{equation}
Linden-Bell (1971) defined that quantity as the 
number of galaxies brighter than $M$
which could have been observed if their magnitude were $M$.
Following Choloniewski (1987), and taking into account the sky coverage of the
2dF present release, the differential LF can be written as
\begin{equation}
\langle \Phi(M) \rangle=\frac{\Gamma \ \sum_i^{M_i\in [M,M+\Delta M]}\psi_i} 
{\Delta M}
\end{equation}
where 
\begin{equation}
\Gamma=\prod_{k=2}^{N}\frac{C_k+w_k}{C_k}\left(V \sum_{i=1}^{N}\psi_i
\sum_{j=1}^{N}\frac{R(\alpha_j,\delta_j)}{N}\right)^{-1},
\end{equation}
\begin{equation}
\psi_k=\prod_{i=1}^{k}\frac{C_i+w_i}{C_{i+1}}
\end{equation}
and $R(\alpha,\delta)$ is the redshift completeness of the parent catalogue.
$C_k \equiv C^{-}(M_k)$ is defined as $C(M)$ but excluding the object
$k$ itself and weighting each object by the inverse of the magnitude-dependent
redshift completeness defined as $w^{-1}=c_z(b_J,\mu)= 0.99 (1-\exp(b_J-\mu))$ 
where $\mu=\mu(\alpha,\delta)$ (see Norberg et al. 2002).

In addition to the non-parametric method described above we also apply
the STY method which uses a maximum likelihood technique to find the most
probable parameters of an analytical $\Phi(M)$, in general assumed to be
a Schechter, function:
\begin{equation}
\Phi(M)\propto 10^{-0.4(M-M^{\ast})(\alpha+1)}\exp{[-10^{-0.4(M-M^{\ast})}]}
\end{equation}
The probability $p_i$ of having an object with absolute magnitude
$M_i$ is 
\begin{equation} 
p_i\equiv \frac{\Phi(M_i)}{\int_{M_{{\rm bright}}}^{M_{{\rm faint}}}\Phi(M)dM}
\end{equation} 
where $M_{\rm bright}$ and $M_{\rm faint}$ are the brightest 
and faintest absolute 
magnitudes observable in the sample at the redshift of the considered galaxy
with the corresponding $k+e$ correction.
Consequently, the method seeks for the parameters $M^{\ast}$ and $\alpha$ 
maximizing the likelihood function
\begin{equation}
{\cal L}=\prod_{i=1}^{N}p_i.
\end{equation}
As we did before, we use the $\mu$-mask to weight each galaxy by 
$1/c_z(b_J,\mu)$
to compensate for the magnitude-dependent incompleteness.
The errors can be estimated from the error ellipsoid defined as 
\begin{equation}
\ln{\cal L}=\ln{\cal L}_{{\rm max}}-\frac12 \chi^{2}_{\beta}(N),
\end{equation}
where $\chi^{2}_{\beta}(N)$ is the $\beta$ point of the $\chi^2$ distribution
with $N$ degrees of freedom.

\begin{figure*}
\epsfxsize=1.0\textwidth
\hspace*{-0.5cm} \centerline{\epsffile{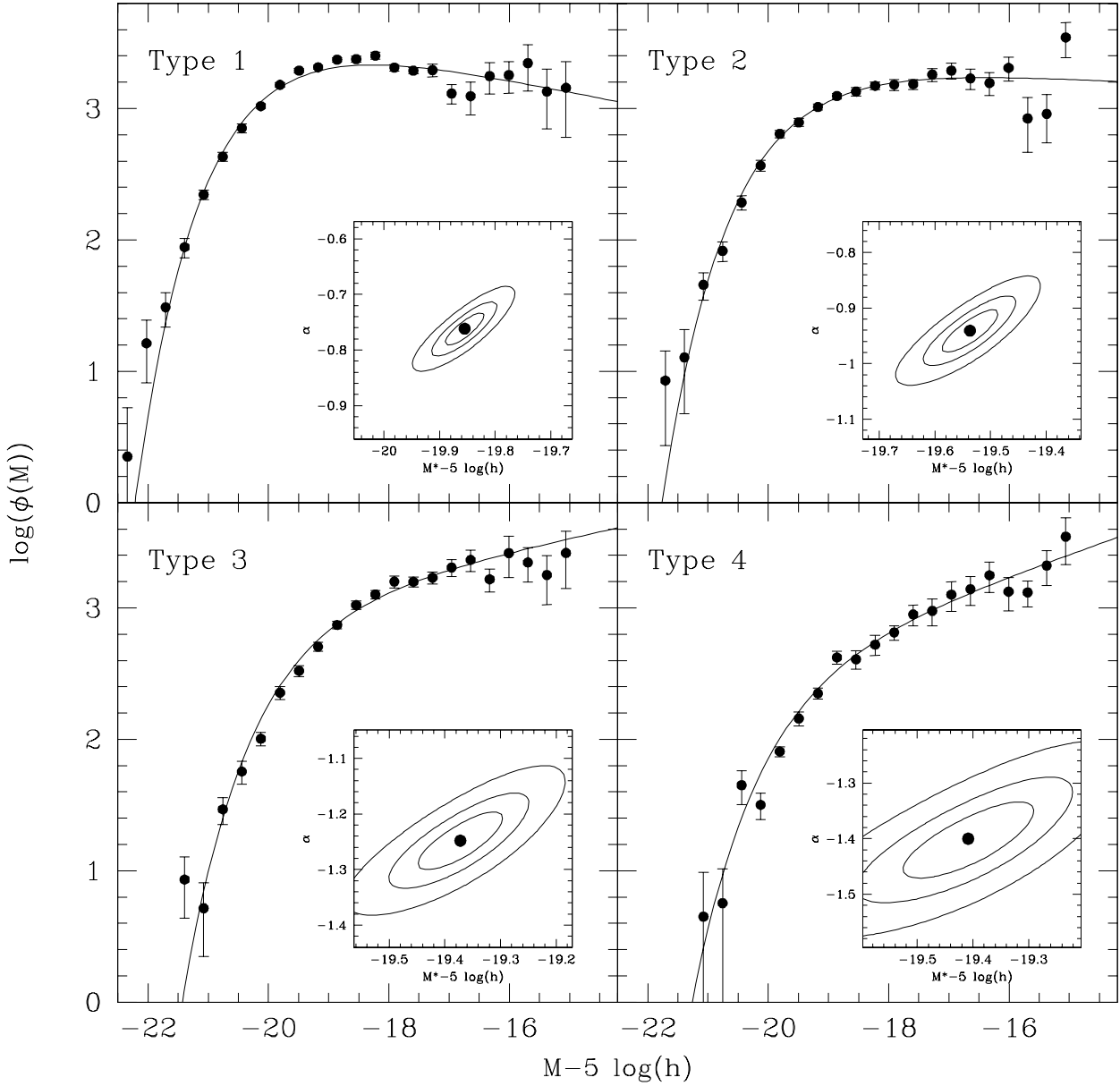}}
\caption{ 
$b_J$ band luminosity function per spectral type for galaxies in groups from the
2dFGGC (arbitrary units) computed using the $C^-$ method.
Error bars are bootstrap resampling technique estimates.
Solid lines show the STY Schechter fits to the data.
Inset panels display $1\sigma, 2\sigma$ and $3\sigma$ confidence ellipses
enclosing the best fit Schechter function parameters, see Table 1. 
}
\label{fig2}
\end{figure*}

The comoving volumes involved in previous equations are estimated
with the general formula
\begin{equation}
V=\frac{\omega c}{H_0} \int_{z_1}^{z_2} \frac{d_L(z)^2(1+z)^{-2}dz}{\sqrt{\Omega_0(1+z)^3+(1-\Omega_0-\Omega_{\Lambda})(1+z)^2+\Omega_{\Lambda}}} 
\end{equation}
where $\omega$ is the solid angle, $c$ is the speed of light, 
$H_0=100 \ h ~\kms \ {\rm Mpc}^{-1}$ and the $d_L(z)$ is the luminosity 
distance. Hereafter we adopt the cosmological model $\Omega_0=0.3$ and
$\Omega_{\Lambda}=0.7$. 

\begin{table*}
\begin{center}
\caption{STY estimates of the Schechter parameters for the luminosity
functions of galaxies in groups}
\begin {tabular}{lcccc}
\hline 
Sample & Redshift range & Number of galaxies & $M^{\ast}-5\log(h)$ & $\alpha$ \\
\hline 
All   & $0.003 - 0.25$ & 14620 & $-19.90\pm 0.03$ & $-1.13\pm 0.02$\\
Type 1 & $0.003 - 0.15$  & 6202 & $-19.85\pm 0.04$ & $-0.76\pm 0.03$\\
Type 2 & $0.003 - 0.15$  & 3445 & $-19.54\pm 0.05$ & $-0.94\pm 0.04$\\
Type 3 & $0.003 - 0.15$ & 1871 & $-19.37\pm 0.08$ & $-1.25\pm 0.05$\\
Type 4 & $0.003 - 0.15$ & 952 &  $-19.40\pm 0.10$ &   $-1.40\pm 0.07$\\
\hline
\end{tabular}
\end{center}
\end{table*}

\subsection{Luminosity function results}
In Figure \ref{fig1} we show the luminosity function for 
all galaxies in groups in
arbitrary units. It should be taken into account that the normalisation 
procedure is only necessary for the group LF. 
Absolute magnitudes are computed using the $k+e$ mean correction for
galaxies in the 2dFGRS as derived by Norberg et al. (2002).
Error bars are estimated using 10 mock catalogues constructed from numerical
simulations of a cold dark matter universe according to the cosmological model 
adopted in this work with a Hubble constant $h=0.7$ and a relative mass 
fluctuation $\sigma_8=0.9$. These simulations were performed using $128^3$
particles in cubic comoving volume of $180 \mpc$ per side. 
The solid line shows the STY best fit which corresponds to the Schechter 
parameters $M^{\ast}=-19.90\pm 0.03$ and $\alpha=-1.13\pm 0.02$. 
Quoted errors are the projections of 1$\sigma$ joint error ellipse (inset plot)
onto each axis. 
The inset plot of Figure \ref{fig1} also shows the error ellipsoids defined by 
2 and 3 $\sigma$ error dispersion.

As pointed out by Mart\'\i nez et al. (2002), the relative fraction of 
the spectral types in groups and field are different, being 
Type 1 galaxies the dominant population in groups.
We are interested in deepening this result by studying the LF for the
four spectral types in groups.
The luminosity functions for different galaxy spectral types are shown in
Figure \ref{fig2}. 
In these computations we applied a redshift cut-off
$z < 0.15$ since $\eta$ determinations are available 
only for galaxies in this redshift range.
In this case, error bars were estimated using the 
bootstrap resampling technique.
The corresponding STY best fit parameters are quoted in Table 1,
and were computed using only those galaxies brighter than 
$M_{b_J}-5\log h=-17$ avoiding points with larger error bars.

At this point, a straightforward comparison between our results
and those obtained for field galaxies in the 2dFGRS by Madgwick et al. (2002)
can be made.
The comparison between the LF Schechter parameters is shown in Figure 
\ref{comp}. Overall, a brightening of the characteristic magnitude is
observed for galaxies in groups with a statistical significance $\sim 2 \sigma$. 
This is particularly significant for 
Type 1 galaxies that are the main contributors to the general LF of
galaxies in groups ($3 \sigma$). 
The denser environment of groups could be thought as the natural
responsible of this brightening. 
In this particular environment, galaxy mergers seems to be the most probable
cause since tidal interactions are more effective when the galaxy
encounter is slow. The later situation, is more frequently observed in
groups where smaller velocity dispersions contrast with the high velocity 
dispersions observed in rich clusters.
Other processes, as ram-pressure (Abadi, Moore \& Bower 1999) or galaxy 
harassment (Moore et al. 1996) should not be as 
effective as in rich cluster where the intra-cluster gas is much denser 
and the interaction rate is higher.\\ 
Nevertheless, the processes mentioned above may be important in the
generation of red low mass galaxies, that are possibly the remnants
of dynamical stripped galaxies in high mass systems 
and can be the responsible for the steepening of the luminosity function 
in clusters.
The observed difference in the $\alpha$ parameter for Type 1 galaxies
(lower panel of Figure \ref{comp})
may be explained in this framework. 
In order to test this scenario, we reanalyze the LF for non star-forming 
galaxies (Type 1) splitting the sample in high ($M \geq 10^{14} \ M_{\odot}$) 
and low ($M < 10^{14} \ M_{\odot}$) group mass subsets.
A significant difference is observed at the faint end slope of the LF from
this comparison, resulting Schechter $\alpha$ parameters of 
($-1.10 \pm 0.06$) and ($-0.71 \pm 0.04$) for high and low mass subsamples
respectively (Figure \ref{type1}). 
These results may support our hypothesis that in groups, mergers 
are probably responsible
for the $M^{\ast}$ brightening, meanwhile process as the ram-pressure
and galaxy harassment could generate the increase at the LF faint end slope 
as observed for non-star forming galaxies in high mass systems. 
Besides the steepening of the faint end slope
of Type 1 galaxies LF in higher mass systems, the galaxy LF for
high and low mass groups do not differ appreciatively between them showing
roughly the same behaviour as the overall LF of galaxies in groups 
(Figure \ref{fig1}). 

Despite the differences between the resulting luminosity functions 
per spectral type, the general LF of galaxies in groups is quite similar 
to that of 2dF field galaxies obtained by Norberg et al. (2002) and 
Madgwick et al. (2002). 
This behaviour could be possibly due to the different 
relative abundances of Type 1 galaxies in the field and groups. 
In groups Type 1 galaxies are by far the dominant population, consequently,
these galaxies almost determine by themselves $M^{\ast}$. Meanwhile, in the
field, Types 2, 3 and 4 contribute more strongly to the LF, generating 
a $M^{\ast}$ brighter than the corresponding to Type 1 galaxies alone.  
Regarding the faint end slope of the overall LF, the main contributors
are late type galaxies (except for high mass groups, as stated above)
in both, field and groups. Since there are no significant differences
between the $\alpha$ parameters for late types in the field and groups,
then the overall faint end slopes are similar.

\section{Luminosity function of groups}
The next step in this work is to analyse the luminosity distribution
of groups independently of the detailed arrangement of luminous material
within each object.
In this section we determine individual group luminosities that allow us
the computation of the group luminosity function and the subsequent
comparison with previous determinations and predictions from semianalytical
models of galaxy formation.  
\subsection{Group luminosities}
We compute group luminosities following the prescription by Moore, Frenk 
\& White (1993). 
For each group the total luminosity, $L_{{\rm tot}}$, is 
the sum of the luminosities of the galaxy members ($L_{{\rm obs}}$) 
plus the integrated 
luminosity of galaxies below the magnitude limit of the survey ($L_{{\rm cor}}$)
\begin{equation}
L_{\rm tot}=L_{\rm obs}+L_{\rm cor}.
\end{equation}
In the computation of $L_{\rm cor}$ we assume that group members are 
independently drawn from the previously computed Schechter fit to the
luminosity function of galaxies in groups. Consequently, the expected 
luminosity of faint group members is
\begin{equation}
L_{\rm cor}=N_{\rm obs}\frac{\int_0^{L_{\rm lim}}L\Phi(L) dL}
{\int_{L_{\rm lim}}^{\infty}\Phi(L) dL},
\end{equation}
where $L_{\rm lim}= 10^{0.4(M_{\odot}-M_{\rm lim})}$, 
$M_{\rm lim}=m_{\rm lim}-25-5\log(d_L(z))$ and $M_{\odot}=5.30$ ($b_J$ band).
The reliability of this scheme for computing luminosities of virialized
systems was widely tested by Moore, Frenk \& White (1993) using groups
from the CfA catalogue and flux and volume limited mock catalogues. 
In order to apply this procedure to our group catalogue we introduce 
some changes that take into account the particular sky coverage of the
parent catalogue: $m_{\rm lim}$ is a function of the angular position of
the group and $N_{\rm obs}=\sum_{i=1}^{N}w_i$, 
where the sum extends over the group members and each member 
is weighted with the inverse of the redshift completeness, 
$w_i=1/R(\alpha_i,\delta_i)$, at its angular position.

\subsection{Group luminosity function}
As a first step in computing the group luminosity function using
the $C^-$ method, it is necessary to find out the completeness limit
in group apparent magnitude of our catalogue.
Rauzy (2001) proposed a new tool in order to define this completeness
limit for a magnitude-redshift sample.
This method does not presuppose that the galaxy population does not 
evolve with time and is homogeneously distributed in space as the 
$V/V_{\rm max}$ test of Schmidt (1968).
Rauzy's method assumes that the sample is complete in apparent magnitude up to
a given magnitude limit $m_{\rm lim}$. The limiting magnitude determines a 
correlation between the absolute magnitude $M$ and the $k+e$ corrected distance modulus 
$Z$.
This method is based on the definition of a random variable
\begin{equation}
\zeta \equiv \frac{\Psi(M)}{\Psi[M_{\rm lim}(Z)]}
\end{equation}
where $M_{\rm lim}(Z)=m_{\rm lim}-Z$. $\zeta$ should be uniformly distributed
between 0 and 1 and should be statistically independent of $Z$ and the 
angular selection function. 
For each object, an unbiased estimate of $\zeta$ is provided by 
\begin{equation}
\hat{\zeta}_i=\frac{r_i}{n_i+1},
\end{equation}
where $r_i$ is the number of objects in the sample with $M \leq M_i$
and $Z \leq Z_i$, and $n_i$ is the number of objects such that $M \leq 
M^i_{\rm lim}(Z_i)$ and $Z \leq Z_i$. 
The expectation $E_i$ and variance $V_i$ of the $\hat{\zeta}_i$ are
$1/2$ and $(n_i-1)/(12(n_i+1))$ respectively.  
Then, the statistic $T_C$, defined as
\begin{equation}
T_C=\frac{\sum_{i=1}^{N_{\rm gal}}\left(\hat{\zeta}_i-\frac12\right)}
{\left(\sum_{i=1}^{N_{\rm gal}} V_i\right)^{1/2}}
\end{equation}
has mean zero and variance unity.
The test consists in computing the quantity $T_C$ on
truncated subsamples according to an
increasing apparent magnitude limit $m_{\ast}$.
While $m_{\ast}$ remains below $m_{\rm lim}$ the subsample is complete 
and $T_C$ statistics is distributed with a mean close to zero and unity 
fluctuations.
When $m_{\ast}$ becomes greater than $m_{\rm lim}$, the 
incompleteness introduces
a lack of objects with $M$ fainter than $M_{\rm lim}(Z)$, therefore $T_C$ 
is expected to fall down systematically to negative values. 
The behaviour of $T_C$ for our
group catalogue is shown in the upper panel of Figure \ref{rauzy}, 
where it can be seen that $T_C$ decreases monotonously 
for limiting apparent magnitudes greater than $m_{\rm lim}\sim 15.6$ 
(vertical solid line),
taking values under $-2$($\equiv -2\sigma_{T_C}$) 
(horizontal long dashed lines).
The lower panel of the same 
figure shows the $k+e$ corrected distance modulus $Z$ as a 
function of the absolute magnitude
for all groups, the vertical solid line corresponds to the $15.6$ 
limiting apparent 
magnitude cutoff as determined with the $T_C$ analysis. 
Dashed horizontal line in the lower panel of Figure \ref{rauzy}
represents a low redshift cut-off ($z_{\rm co}=0.04$)
that we impose in order to avoid small volume undersampling effects.

\begin{figure}
\epsfxsize=0.5\textwidth
\hspace*{-0.5cm} \centerline{\epsffile{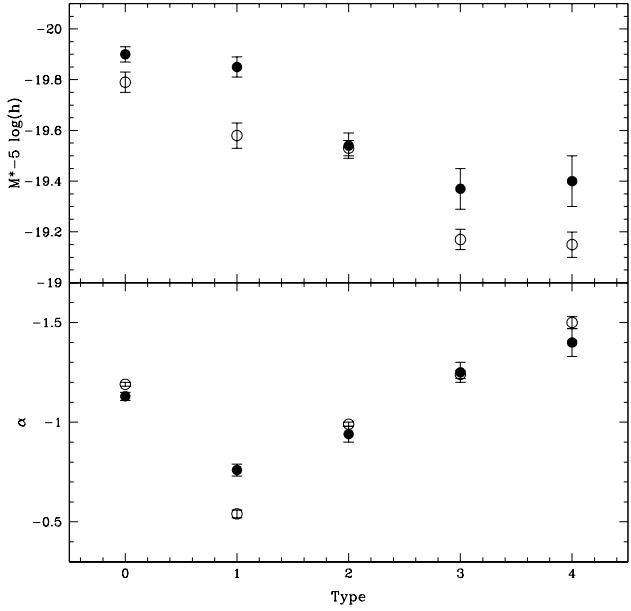}}
\caption{Comparison between the LF Schechter parameters of field
galaxies by Madgwick et al. (2002) (open circles) 
and galaxies in groups of this work (filled circles). 
The upper panel compares the characteristic magnitude $M^{\ast}-5\log(h)$,
 meanwhile
the lower panel is the comparison for the faint-end slope $\alpha$.
Both panels are plotted as a function of galaxy spectral type.
We label as Type 0 all galaxies, irrespectively of spectral type.
}
\label{comp}
\end{figure}

\begin{figure}
\epsfxsize=0.5\textwidth
\hspace*{-0.5cm} \centerline{\epsffile{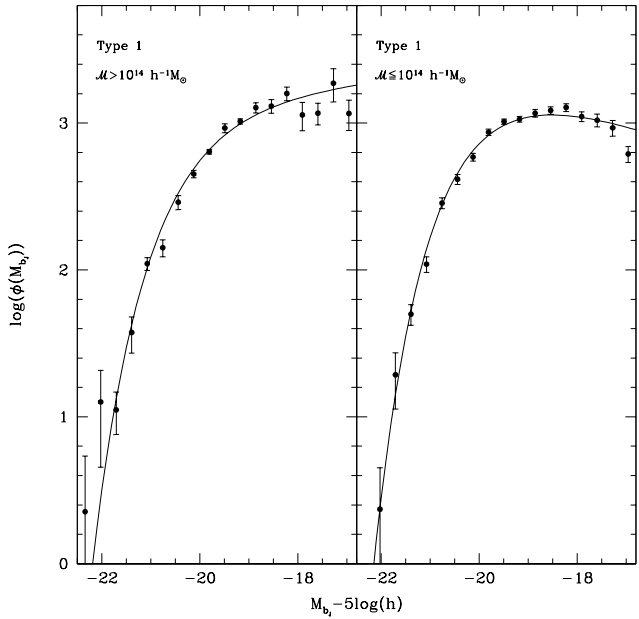}}
\caption{
Luminosity functions for Type 1 galaxies in high (left panel) and low
(right panel) mass groups samples.
Solid lines are the best Schechter fit to the luminosity function
determined with the STY method.
}
\label{type1}
\end{figure}

Finally, our flux-limited group sample comprises 922 groups with the 
constraints $0.04\leq z\leq 0.25$ and $m_{b_J}\leq 15.6$.
We compute for this sample the group LF with the $C^-$ method in a 
similar way as we did for galaxies in groups in Section 3, but
in this case no weight is needed in the $C_k\equiv C^-(M_k)$ computation 
(i.e. $w \equiv 1$ in equations 4 and 5) since the choice of the magnitude 
cut-off ensures the completeness of the final sample.
The resulting group LF is shown in Figure \ref{groupLF}.
Error bars were computed by applying the same LF estimator to the mock
catalogues in a similar way as explained is Section 3.
In the same figure we show the results from groups in the CfA redshift survey
by Moore, Frenk \& White (1993) and from semianalytical models by 
Benson et al. (2000) for a $\Lambda$ cold dark matter model. 
A general good agreement with previous results
is observed except for $M_{b_J}-5\log h> -21.5$. Since the 2dFGGC 
has only groups with at least 4 members, there is a lack of low luminosity
systems that could explain the differences in the low luminosity tail
in Figure \ref{groupLF} when comparing with previous works.  
As shown by Moore, Frenk \& White (1993), the faint end
of the group luminosity function is made up almost enterely of
single galaxies; binaries and groups with three members link this
to the steep bright tail of richer groups and clusters.
We have checked whether the 4-member criterion is affecting the
faint-end of the group luminosity function by re-computing it including
groups with 3 and 2 members. The resulting group LF is in agreement with the
results of Moore, Frenk \& White (1993) in the whole absolute magnitude range.
Nevertheless, it should be recalled that the 4 member cut-off 
is necessary in order to avoid spurious detections.
As stated by Merch\'an \& Zandivarez (2002),
this false detections can reach aproximately $\sim 40\%$ when considering
groups with only three members.  
Consequently, the 4-member criterion ensures a reliable determination 
of the luminosity
function of virialised systems over a wide range of luminosities.

\begin{figure}
\epsfxsize=0.5\textwidth
\hspace*{-0.5cm} \centerline{\epsffile{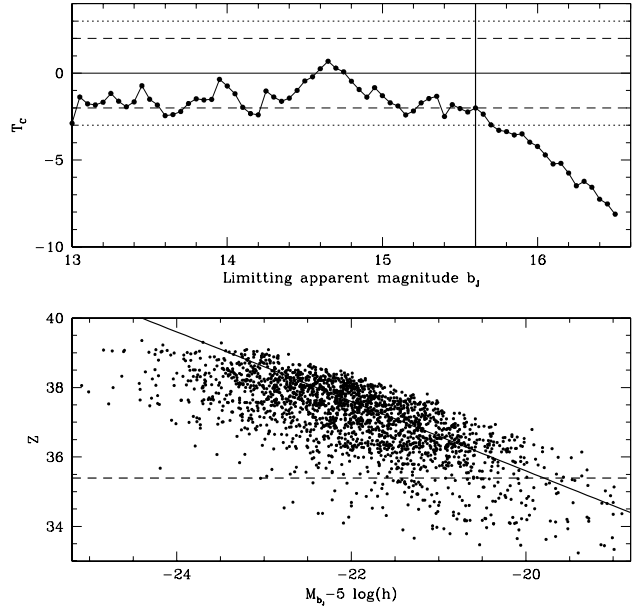}}
\caption{ 
Upper panel:
the test for completeness $T_C$ (Rauzy 2001) applied to 2dFGGC.
Apparent $b_J$ magnitudes were computed using group total luminosities
following Moore, Frenk \& White (1993) and group redshifts.
A systematic decline of the $T_C$ statistics can be observed
beyond $m_{\rm lim}=15.6$ (vertical line). Lower panel:
Distance modulus $Z$ versus absolute magnitude for groups.
Solid line shows the apparent magnitude of completeness $b_J=15.6$.
Horizontal dashed line represents the low redshift cut-off $z>0.04$. 
}
\label{rauzy}
\end{figure}

\begin{figure}
\epsfxsize=0.5\textwidth
\hspace*{-0.5cm} \centerline{\epsffile{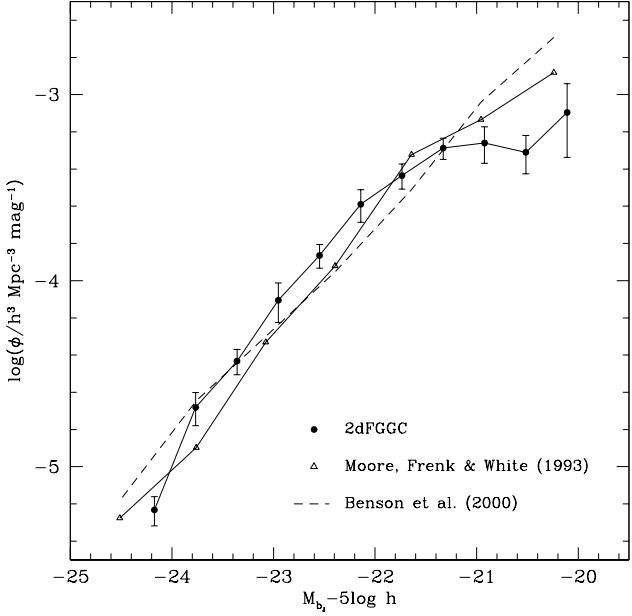}}
\caption{ 
Group luminosity function (solid circles) for our flux-limited
group sample compared with the CfA redshifts survey groups (open
triangles) Moore, Frenk \& White (1993) and the semianalytical models
of Benson et al (2000) (dashed lines). 
}
\label{groupLF}
\end{figure}

\section{Mass function of groups}
The statistical strength of our group sample allows us an accurate measurement
of the group mass function. 
In this section we explore the mass function of intermediate mass systems,
using the flux-limited sample defined in the previous section.
For this purpose we use the $1/V_{\rm max}$ procedure, that is a non-parametric
method consisting in weight each object with the inverse of the maximum
comoving volume of the survey, $V_{\rm max}$, 
\begin{equation}
V_{\rm max}(M_i)=\int_{{\rm{min}}(z_2,z_{\rm max})}^{{\rm{max}}
(z_1,z_{\rm min})}
\frac{dV}{dz}dz
\end{equation}
in which it remains observable given all the observational selection limits. 
The extremes of the volume integral are found by solving the equations
\begin{equation}
M_i=\left\{ 
       \begin{array}{l}
        m_2-5\log d_L(z_{\rm max})-25-k(z_{\rm max}) \\
        m_1-5\log d_L(z_{\rm min})-25-k(z_{\rm min}) 
       \end{array}
    \right.
\end{equation}
that are the redshifts of objects with the same absolute magnitude, $M_i$, 
of the considered object, but with apparent magnitude at the bright and faint 
limits of the survey.
Consequently, the differential mass function can be computed as
\begin{equation}
\frac{dN}{d{\mathcal M}}({\mathcal M})= \sum_{{\mathcal |M_{\it i}-M|}\leq 
d{\mathcal M}/2}\frac{1}{V_{\rm max}(M_i)\hat{R}(\alpha_i,\delta_i)}
\end{equation}
where ${\mathcal M}_i$ are the group masses and 
$\hat{R}(\alpha_i,\delta_i)$ are the mean group redshift completeness.

For a meaningful comparison between the resulting group mass function and 
analytical mass function predictions (Press \& Schechter 1974,
Sheth \& Tormen 1999 and Jenkins et al. 2001), it is necessary to relate 
the group virial masses with the mass definition in the models.
Group virial masses were computed inside a galaxy density
contrast $(\delta \rho/\rho)_g=80$ (Merch\'an \& Zandivarez 2002), 
which corresponds to a mass 
density contrast $(\delta \rho/\rho)_m=(\delta \rho/\rho)_g/b$,
where $b$ is the linear bias factor.
If we estimate $b=1/\sigma_8$, assuming $\sigma_8=0.9$, the resulting
density contrast is $(\delta \rho/\rho)_m=72$. 
In order to make a comparison with the 
analytical predictions, the masses should be computed for a particular
density contrast depending on the cosmological model.
Taken into account the relation between the virial density of collapsed
objects in units of the critical density and the adimensional
density parameter $\Omega_0$, as quoted by Eke, Cole \& Frenk (1996), 
we must have $(\delta \rho/\rho)_m \sim 330$ for our cosmological model,
assuming spherical collapse approximation.      
To find the relation between the group virial masses and that corresponding 
to the matter overdensity $330$, we assume a simple mean overdensity profile 
for groups that scales as $r^{-2}$, where $r$ is the group-centric distance.
Given that in this case ${\mathcal M} \propto r$ then the 
resulting scaling relation is
${\mathcal M}_{330} \simeq {\mathcal M}_{72} \sqrt{72/330}$.

\begin{figure}
\epsfxsize=0.5\textwidth
\hspace*{-0.5cm} \centerline{\epsffile{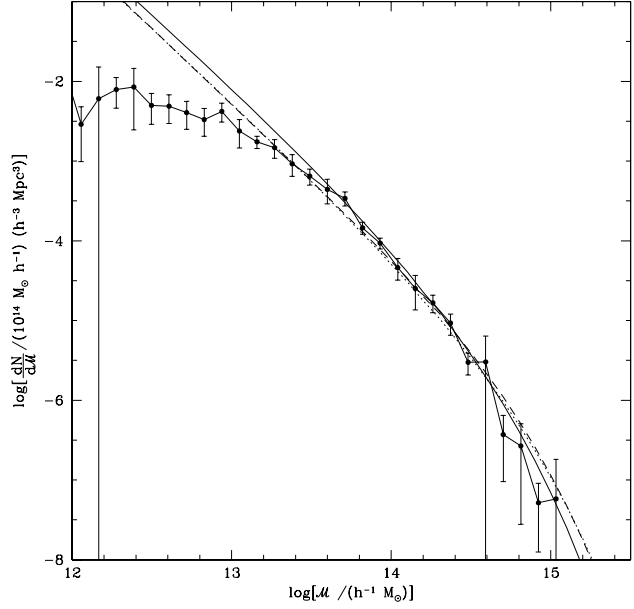}}
\caption{Group mass function for the 2dFGGC (dots). Error bars correspond
to one $\sigma$ dispersion obtained from ten mock catalogues.   
The solid line, dotted line and dashed line are the Press \& Schechter (1974), 
Sheth \& Tormen (1999) and Jenkins et al. (2001) prescriptions respectively.}
\label{fig6}
\end{figure}

Figure \ref{fig6} shows the comparison between the group mass function, with
masses scaled using the previous relation, and the analytical prediction
for Press \& Schechter (1974), Sheth \& Tormen (1999) and Jenkins et al. (2001).
Overall, analytical predictions are in excellent agreement with our 
results spanning two orders of magnitude in mass 
$(10^{13}h^{-1} M_{\odot} \lsim {\mathcal M} \lsim 10^{15} h^{-1} M_{\odot})$.
The differences for masses smaller than $\sim 10^{13} h^{-1} M_{\odot}$
arise as consequence of the lack of low mass groups, mainly due to the 
member number cut-off imposed to the 2dFGGC.

\section{CONCLUSIONS}
Here, we have applied several statistical analysis to groups
and galaxies in groups taken from  an updated version of the 
Merch\'an \& Zandivarez (2002) group catalogue (2dFGGC).
We have focused on an accurate determination of the luminosity functions
of galaxies in groups, and the luminosity and mass functions
of groups taking advantage of the statistical power of the 2dFGGC.

In the LF computations,
we have used a version of the Choloniewski (1987) approach to the
$C^-$ method by Lynden-Bell (1971), adapted to the particular sky coverage
of the 2dFGRS. 
The choice of this particular estimator of the luminosity function
is inspired in the conclusions of the comparative analysis of LF estimators
by Willmer (1997), who states that the $C^-$ method is less affected
by inhomogeneities in the sample.
The resulting luminosity function for galaxies in groups (Figure \ref{fig1})
is well fitted by a Schechter function with shape parameters
$M^{\ast}-5\log(h)=-19.90\pm 0.03$ and $\alpha=-1.13\pm 0.02$ as 
determined by a STY fitting procedure.
These values are quite consistent with those obtained by 
Norberg et al. (2002) for field galaxies in the 2dFGRS.

We have performed a similar analysis in subsamples of galaxies 
(Figure \ref{fig2} and Table 1) defined
by the spectral types of Madgwick et al. (2002).
In general, the characteristic magnitudes $M^{\ast}$ are shifted
to higher luminosities with respect to the field values
found by Madgwick et al (2002), irrespectively of spectral type.
This shift may be due to galaxy dynamical interactions such as mergers,
which are expected to be much more frequent in systems with low
velocity dispersions as the majority of our group sample.

The faint end slopes of the luminosity functions, $\alpha$, are
consistent with those corresponding to field galaxies except for
low star forming, Type 1, galaxies, that show a steeper value in groups
(Figure \ref{comp}).
We deepen our analysis for Type 1 galaxies exploring the behaviour of
$\alpha$ for two subsets of groups: low 
(${\cal M}\leq 10^{14} h^{-1} M_{\odot}$) and high
(${\cal M}>10^{14} h^{-1} M_{\odot}$), virial masses.
We observe an increase in the faint end slope
of the Type 1 LF ($\alpha=-1.10\pm 0.06$) for galaxies in high mass systems
meanwhile for galaxies in low mass systems it remains closer to the 
global value ($\alpha=-0.71\pm 0.04$) (Figure \ref{type1}).
This effect could be the result of internal processes in higher mass
system environments such as ram-pressure and galaxy harassment, that
are not expected to be significantly important in smaller overdensities.

We have defined group luminosities following Moore, Frenk \& White (1993),
as the sum of its observed luminosity plus a normalised integral
of the galaxy luminosity function below the flux limit of the survey.
This definition has proved to be reliable in the computation of group 
total luminosity using observations and artificial flux and volume-limited
catalogues. 
The main aim of this computation is the determination of the luminosity
function for our group catalogue. 
Since the $C^{-}$ method for luminosity function estimation requires
a fair selection criterion, we have used the Rauzy (2001) $T_C$ statistics
to determine an apparent magnitude cut-off for the 2dFGGC that
ensures the highest level of completeness.
Our final flux-limited group sample comprises 922 groups with
apparent magnitudes brighter than $b_J=15.6$ (Figure \ref{rauzy}).
The resulting group luminosity function for this sample (Figure \ref{groupLF}) 
is consistent with Moore, Frenk \& White (1993) determination for groups in
the CfA redshift survey and with the semianalytical model prediction 
of a $\Lambda$ cold dark matter cosmology performed by Benson et al. (2000).
This agreement is acceptable in the luminosity range $M_{b_J}\lsim -21.5$. 
The differences observed at fainter luminosities are mainly due to
the lack of groups with less than 4 members in the 2dFGGC.
The intrinsic characteristics of the group finding algorithm
used in the construction of the 2dFGGC by Merch\'an \& Zandivarez (2002),
determine that below this limit any resulting sample has an unacceptable
level of contamination by spurious detections.

The flux-limited group sample adopted in the luminosity 
function computation is also fair enough for the determination
of the group mass function.
This determination was achieved by using an adapted version of the
$1/V_{\rm max}$ that considers the sky coverage of the group sample. 
Finally, a comparison with analytical predictions of halo abundances
is made using a simple scheme to relate group virial masses and
that corresponding to the adequate overdensity in our cosmological model.
The results are displayed in Figure \ref{fig6}, 
where it can be seen a notorious
agreement with the Press \& Schechter (1974), Sheth \& Tormen (1999)
and Jenkins et al. (2001) mass functions for masses ${\mathcal {M}}
\gsim 10^{13} h^{-1}M_{\odot}$. Again, the disagreement for low masses
can be attributed to absence of poor groups.

The statistical significance of the group catalogue used in this work
allowed us to obtain very important clues about internal properties 
of intermediate mass systems and also to observe the level of agreement
obtained from both analytical and semianalytical predictions in the
framework of a $\Lambda$ cold dark matter model.

\section*{Acknowledgments}
We are indebted with Diego G. Lambas and Hern\'an Muriel
for interesting comments and discussions.
We thank to Peder Norberg and Shaun Cole for kindly providing the 
software describing the mask of the 2dFGRS and to the 2dFGRS Team
for having made available the current data sets of the sample.
We would like to thank J. Coltrane.
HJM, AZ and MEM are supported by fellowships from CONICET, Argentina.
MJLD is supported by a fellowship from SECyT, Universidad Nacional de C\'ordoba,
Argentina.
This work has been partially supported by grants from 
Consejo de Investigaciones 
Cient\'{\i}ficas y T\'ecnicas de la Rep\'ublica Argentina (CONICET), the
Secretar\'{\i}a de Ciencia y T\'ecnica de la Universidad Nacional de C\'ordoba
(SeCyT), Agencia Nacional de Promoci\'on Cient\'\i fica de la Rep\'ublica
Argentina and Agencia C\'ordoba Ciencia.

\end{document}